\documentclass[]{jfm}
\usepackage{graphicx}
\usepackage{newtxtext}
\usepackage{newtxmath}
\usepackage{natbib}
\usepackage{hyperref}
\hypersetup{
    colorlinks = true,
    urlcolor   = blue,
    citecolor  = black,
}

\newcommand{\RomanNumeralCaps}[1]
\linenumbers

\usepackage{graphicx}
\usepackage{epstopdf, epsfig}

\shorttitle{Hanging Droplets from Liquid Interfaces}
\shortauthor{P. Singh, A. Pal and N. Singh}

\title{Hanging Droplets from Liquid Interfaces}

\author{Piyush Singh,
  Anikesh Pal
  \corresp{\email{pala@iitk.ac.in}}
 \and Narinder Singh}

\affiliation{Department of Mechanical Engineering, Indian Institute of Technology,
Kanpur 208016, India}

\begin{document}

\maketitle

\begin{abstract}
The impact of a heavier droplet into a deep pool of lighter liquid is investigated using three-dimensional numerical simulations. Unprecedented to any numerical simulations, we demonstrate that the heavier droplets can hang from the surface of a lighter liquid using surface tension. The impact phenomenon and the evolution of the heavier droplet as a function of its size and release height are explored.  A theoretical model is also formulated to understand the role of different forms of energies associated with the hanging droplet. We further solve the force balance equations for the hanging droplets analytically and demonstrate that the results obtained from our simulations match very well with the analytical solution. This research offers opportunities in many areas, including drug and gene delivery, encapsulation of biomolecules, microfluidics, soft robots, and remediation of oil spills.
\end{abstract}

 
\section{Introduction}
Small living creatures such as water striders, beetle, and mosquito larvae use surface tension to stand, walk, leap, or hang on the surface of water \citep{bush2006walking,feng2007superior,hu2005meniscus,vella2015floating,lee2017floating}. Inspired by these natural occurrences, researchers have developed millimeter-scale robots \citep{koh2015jumping,hu2018small} for transport across the surface of a liquid that might be useful in targeted drug delivery, minimal invasive surgery, and other bio-engineering applications. These robots feature a hydrophobic surface with strong interfacial tension that prevents the body from breaking the liquid surface and sinking. Once the body rests at the surface, additional locomotion can be provided utilizing the techniques described by \cite{hu2018small, jiang2019directional, grosjean2018magnetocapillary}. Many other biomedical applications require encapsulation of one liquid in another. Examples include separation \citep{peters1987partition, zhang2016one,li2018light} or encapsulation \citep{delcea2011stimuli,orive2003cell} of bio-molecules and cells. In this context, aqueous two-phase systems (ATPSs) \citep{hann2016one,hann2017awe,chao2018generation,xie2019compartmentalized}, formed using a mixture of dextran and poly(ethylene glycol) (PEG) which phase separates to form two immiscible aqueous phases, are widely used. \\

We perform the first-of-its-kind three-dimensional numerical simulations on two immiscible aqueous solutions of dextran and PEG to demonstrate that a droplet of higher density (dextran) can either hang from the surface like mosquito larvae, bounce on the surface like water striders or form a shroud that completely wraps the denser fluid as it sinks in the pool of a lighter liquid (PEG solution). As the drop makes contact with the pool, the evolution of the three-phase contact line (TPCL) plays a major role in the dynamics of a drop hanging or sinking from the surface. It will be shown using force balance equations that during the hanging process the surface tension force balances the heavier droplet at the surface of the pool. The size of the droplet and its initial kinetic energy are some of the key parameters that dictate the outcome in this situation. \cite{xie2020} experimentally presented a similar phenomenon of hanging \citep{phan2012can,phan2014stability} and wrapping \citep{kumar2018wrapping} using ATPS of a dextran solution containing polycations and PEG solution containing polyanions. In the presence of oppositely charged polyelectrolytes the solutions after coming into contact create structured coacervate sacs of negligible mass and thickness at their interface. These coacervate sacs effectively increase the interfacial tension between the two solutions resulting in hanging of the heavier droplets from the pool surface. \\

\section{\label{sec:appA}Methods}
The volume of fluid (VOF) approach of \cite{hirt1981volume} serves as a foundation for calculations involving two fluids separated by a sharp interface. The VOF approach achieves excellent compliance with mass conservation, but it can be difficult to capture the geometric features of a complex interface. \cite{osher1988fronts} introduced the level set (LS) method, which is an efficient interface capture technique. This approach properly captures the interface, although it may violate mass conservation in some circumstances. A combination of the LS approach with the VOF method, known as the coupled level set and volume of fluid (CLSVOF) method can accomplish mass conservation and properly capture the interface. The LS function is utilized exclusively to compute the geometric characteristics at the interface in the CLSVOF technique \citep{sussman2000coupled}, while the volume fraction is determined using the VOF method. Continuum surface tension force (CSF) by \cite{brackbill1992continuum} has been widely used to evaluate the source term due to surface tension. However a free energy-based surface tension force (FESF) model is proposed by \cite{yuan2017free} for simulation of multi-phase flows by level set method, which outperforms the previous CSF model in terms of accuracy, stability, convergence speed and mass conservation. \cite{howard2021conservative} also extended the conservative LS method for N fluid phases by introducing a new compression-diffusion equation which handles large deformation and triple junctions more accurately. In order to solve the N-phase flow problems, algorithms with $N$ \citep{ruuth1998diffusion}, $N-1$ \citep{smith2002projection, zlotnik2009hierarchical}, $N(N-1)/2$ \citep{starinshak2014new, starinshak2014new1} and $\log_2N$ \citep{chan2001active} LS functions have been used.\\

\subsection{Governing Equations}

Considering incompressible Newtonian fluids, the mass and momentum conservation equation for fluids 1, 2 and 3 are given by
\begin{equation}
    \label{continuity}
    \nabla\cdot\mathbf{U} = 0,
\end{equation}
%\begin{widetext}
\begin{equation}
    \label{momentum}
    \rho\left(\frac{\partial \mathbf{U}}{\partial t} + \mathbf{U}\cdot\nabla\mathbf{U}\right) = -\nabla\mathcal{P} + \nabla\cdot\left(2\mu\mathbf{D} \right) + \mathbf{F} + \mathbf{F_{st}}.
\end{equation}

%\end{widetext}
where, $\mathbf{U}$ is the velocity vector field with components ($\mathrm{U}_1, \mathrm{U}_2, \mathrm{U}_3$), $\mathcal{P}$ represents the dynamic pressure, $\rho$ and $\mu$ are scalar fields representing density and dynamic viscosity, $\mathbf{D}$ is the deformation tensor, and $\mathbf{F}$ and $\mathbf{F_{st}}$ are body force and surface tension force per unit volume.\\

\begin{equation}
    \label{deform_tensor}
    \mathbf{D} = \frac{1}{2}\left( \nabla U + \nabla U^T\right).
    % \mathrm{D_{ij}} = \frac{1}{2}\left(\frac{\partial u_i}{\partial u_j} + \frac{\partial u_j}{\partial u_i} \right).
\end{equation}

Gravitational force is the only body force acting on all the fluids in our case. Surface tension force, as given by \cite{howard2021conservative}, is used in the momentum equation as follows:\\

\begin{equation}
%\begin{multline}
    \label{momentum2}
    \rho\left(\frac{\partial \mathbf{U}}{\partial t} + \mathbf{U}\cdot\nabla\mathbf{U}\right) = \\-\nabla\mathcal{P} + \nabla\cdot\left(2\mu\mathbf{D} \right) + \rho\mathbf{g} + \sum_{i,j=1}^{3} \nabla\cdot\left(\frac{3}{2}\sigma_{ij}\varepsilon\nabla\varphi_i\times\nabla\varphi_j\right).
\end{equation}

%\end{multline}
Here, $\mathbf{g}$ is the acceleration due to gravity. $\sigma_{ij}$ represents the surface tension at the interface between the fluids $i$ and $j$. Interfacial numerical thickness, $\varepsilon$ is defined based on grid size, $\Delta x$ as $\varepsilon=k_\varepsilon\Delta x$; for all the simulations reported here, $k_\varepsilon=1.5$ is used. The surface tension force is obtained using the free energy surface tension force model. The free energy density for N immiscible fluids is given by \cite{DONG2014691}. \\

In the present work, CLSVOF is used which combine the advantages of both level set method and volume of fluid method. The LS function is defined as a signed distance function from the phase interface such that:
\begin{equation}
    \label{LS_def}
    \phi_i(\mathbf{x}) \left\{
    \begin{array}{cl}
      > 0 & \mbox{if }\mathbf{x}\in\Omega_i \\[2pt]
      = 0 & \mbox{if }\mathbf{x}\in\Gamma_i \\[2pt]
      < 0         & \mbox{if }\mathbf{x}\notin\Omega_i,
    \end{array} \right.
\end{equation}
where $\Omega_i$ denotes the subdomain containing the fluid of the $i$th phase and $\Gamma_i$ is the sharp interface of the $i$th phase. The VOF function is taken as the fraction grid cell volume occupied by fluid of phase $i$. The VOF function is defined so as to ensure the following condition
\begin{equation}
    \label{VOF_sum}
    \sum_{i=1}^{3}f_i = 1.
\end{equation}
The scalar field $\varphi_i$ used in \ref{momentum2} is defined using the LS function as:
\begin{equation}
    \label{colorfnc}
    \varphi_i = \mathrm{H}(\phi_i).
\end{equation}
Here Heaviside function is defined as follows:\\
\begin{equation}
    \label{hvs}
    \mathrm{H}(\phi) = \left\{ \begin{array}{cc} 
                0 & \hspace{5mm} \mbox{if} \hspace{2mm} \phi < -  \varepsilon  \\
               \frac{1}{2}\left [ 1+\frac{\phi }{\varepsilon} +\frac{1}{\pi }\sin \left (\frac{\pi \phi}{\varepsilon }   \right ) \right ] & \hspace{5mm} \mbox{if} \hspace{2mm} \left | \phi  \right | = \varepsilon \\
                1 & \hspace{5mm} \mbox{if} \hspace{2mm} \phi > + \varepsilon. \\
                \end{array} \right.
\end{equation}
The varying density and viscosity fields are also defined using the Heaviside function as:
\begin{equation}
    \label{density}
    \rho = \rho_1\mathrm{H}(\phi_1)+\rho_2\mathrm{H}(\phi_2)+\rho_3[1-\mathrm{H}(\phi_1)-\mathrm{H}(\phi_2)],
\end{equation}
\begin{equation}
    \label{viscosity}
    \mu = \mu_1\mathrm{H}(\phi_1)+\mu_2\mathrm{H}(\phi_2)+\mu_3[1-\mathrm{H}(\phi_1)-\mathrm{H}(\phi_2)].
\end{equation}

The motion of interfaces is tracked by explicitly solving the advection equation for both LS and VOF functions.\\
\begin{equation}
    \label{LS}
    \frac{\partial \phi_i}{\partial t} + \nabla \cdot (\mathbf{U}\phi_i) = 0,
\end{equation}
\begin{equation}
    \label{VOF}
    \frac{\partial f_i}{\partial t} + \nabla \cdot (\mathbf{U} f_i) = 0.
\end{equation}
%The equation for advection of VOF function gives better mass conservation which LS function would not give alone. The LS function, on the other hand, ensures sharp interface boundaries, which would have diffused if only VOF was used.
\subsection{Boundary Conditions}
The governing equations are solved in a three-dimensional cartesian space. A closed system is considered for the simulations such that no fluid enters or leaves the computational domain.
\begin{equation}
    \label{impermeable}
    \mathbf{U}\cdot\mathbf{n} = 0.
\end{equation}
No-slip boundary condition is assumed at all the boundaries of the computational domain. The boundaries of the computational domain are kept sufficiently away from the droplet to ensure that it does not affect the dynamics of the flow.
\begin{equation}
    \label{noslip}
    \mathbf{n}\times\mathbf{U} = 0.
\end{equation}
Therefore, a dirichlet boundary condition is used for the velocity field on all the boundaries. On the other hand a neumann boundary condition is used for pressure at the boundaries.

\begin{equation}
    \label{pressurebc}
    \mathbf{n}\cdot\nabla\mathcal{P} = 0.
\end{equation}

\subsection{Numerical Methods}
The governing partial differential equations are advanced in time using an explicit 3rd order Runge-Kutta method \citep{williamson1980low}. %The PDEs are decomposed into ODEs by explicit treatment of the terms. 
A staggered grid is used for the discretization in space where the vector field quantities (like $\mathbf{U}$) are defined at the cell face center and the scalar field quantities ($\mathcal{P},\rho,\mu,\phi,f$) are defined at the cell center. The advection terms are discretized using a second-order ENO scheme as used by \cite{chang1996level} and \cite{son2007level}. The viscous terms are discretized using a second-order central difference scheme. It is to note here that the viscosity, $\mu$, is not constant throughout the domain and so special care has to be taken to include $\mu$ into the discretization scheme. The pressure Poisson equation, which is employed to project a velocity field into a divergence-free space, is solved using a parallel multigrid iterative solver \citep{pal2020evolution, pal2020deep} to obtain the dynamic pressure. To advance in time for the advection equations of the LS and the VOF functions, we use an operator splitting algorithm \citep{son2003efficient}, in which we solve equation \ref{LS} and equation \ref{VOF} one direction at a time. The operator splitting is of second-order accuracy in time and the order of sweep direction at each time step is also alternated. The solution of the advection equation for the LS function does not satisfy the signed distance property from the interface. For this the LS function is reinitialized at each time step after operator the splitting algorithm \citep{son2003efficient}.\\

To perform a three-phase flow simulation ($\mathrm{N}=3$), only two ($\mathrm{N}-1$) phase equations are solved using the CLSVOF algorithm. The numerical solution to these $\mathrm{N}-1$ phase equations generates some voids and overlaps between the phases. By using the $\mathrm{N}-1$ phase equation, it is assumed that the $\mathrm{N}$th phase occupies the void region and this avoids the singularity problems in the computational domain.
\begin{equation}
    f_3 = 1-f_1-f_2,
\end{equation}
\begin{equation}
    \mathrm{H}(\phi_3)=1-\mathrm{H}(\phi_1)-\mathrm{H}(\phi_2).
\end{equation}
A VOF correction is performed to overcome the overlap issues such that
\begin{equation}
    \label{VOF_correction}
    f_2 = 1 - f_1, \qquad \mbox{if } f_1+f_2> 1.
\end{equation}
This VOF correction is biased toward phase 1 as we assume phase 1 to represent the primary fluid of interest. The coupled nature of the CLSVOF algorithm appropriately adjusts the LS function for this VOF correction.\\

A constant time step size is used such that it satisfies the following time step restrictions. Firstly, the standard Courant-Friedrichs-Lewy (CFL) condition is satisfied.
\begin{equation}
    \label{CFL}
    \Delta t_u \le \frac{\mathrm{CFL}}{\frac{|u|_{\mathrm{max}}}{\Delta x} + \frac{|v|_{\mathrm{max}}}{\Delta y} + \frac{|w|_{\mathrm{max}}}{\Delta z}}.
\end{equation}
According to \cite{brackbill1992continuum} when treating the surface tension term explicitly, the time step must be sufficiently small to resolve the capillary waves phenomena. This gives another time step restriction as:\\

\begin{equation}
    \label{capilary_time_restriction}
    \Delta t_\sigma \le \mathrm{CFL}_\sigma \sqrt{\frac{\mathrm{min}(\rho_i+\rho_j)\times\mathrm{min}(\Delta x,\Delta y, \Delta z)^3}{\mathrm{max}(4\pi\sigma_{ij})}}, \qquad i\ne j.
\end{equation}

We have used $\mathrm{CFL}=0.5$ and $\mathrm{CFL}_\sigma = 0.5$ for all cases. Other time step restriction criteria based on viscosity and gravity give more relaxed values. A constant time step is used such that it satisfies the above-mentioned restrictions sufficiently throughout the simulation.\\

\section{Validation of Numerical Approach}
\label{result}

\subsection{Advection Test}
In this study, a parallel three-phase incompressible flow solver is used which is an extension of an existing two-phase flow solver. Hence an advection test on a two-phase flow solver using parallel computations was performed first. The flow domain $\Omega$ is a cube of length $1$ and a sphere of radius $0.15$ is placed at $\mathbf{x}=(0.35,0.35,0.35)$. A 3D shear deformation field is defined as:
\begin{equation}
    \mathrm{U}_1 =2\sin^2(\pi x)\sin(2\pi y)\sin(2\pi z)\cos(\frac{\pi t}{T}),
\end{equation}
\begin{equation}
    \mathrm{U}_2 =-\sin^2(\pi y)\sin(2\pi x)\sin(2\pi z)\cos(\frac{\pi t}{T}),
\end{equation}
\begin{equation}
    \mathrm{U}_3 =-\sin^2(\pi z)\sin(2\pi x)\sin(2\pi y)\cos(\frac{\pi t}{T}),
\end{equation}
with time $t$ and time period $T$. The $\cos(\frac{\pi t}{T})$ term makes the velocity field periodic with respect to time and ensure that the time integral over a time period at any point in the domain results in zero. This means that any particle moving in the domain will return to its initial position after one time period. Hence it is expected that the sphere will deform under the shear velocity field, get stretched and then eventually return to its initial shape and position.\\

\begin{figure*}
\centering
\includegraphics[width=1.0\textwidth,trim={0cm 0cm 0 0cm},clip] {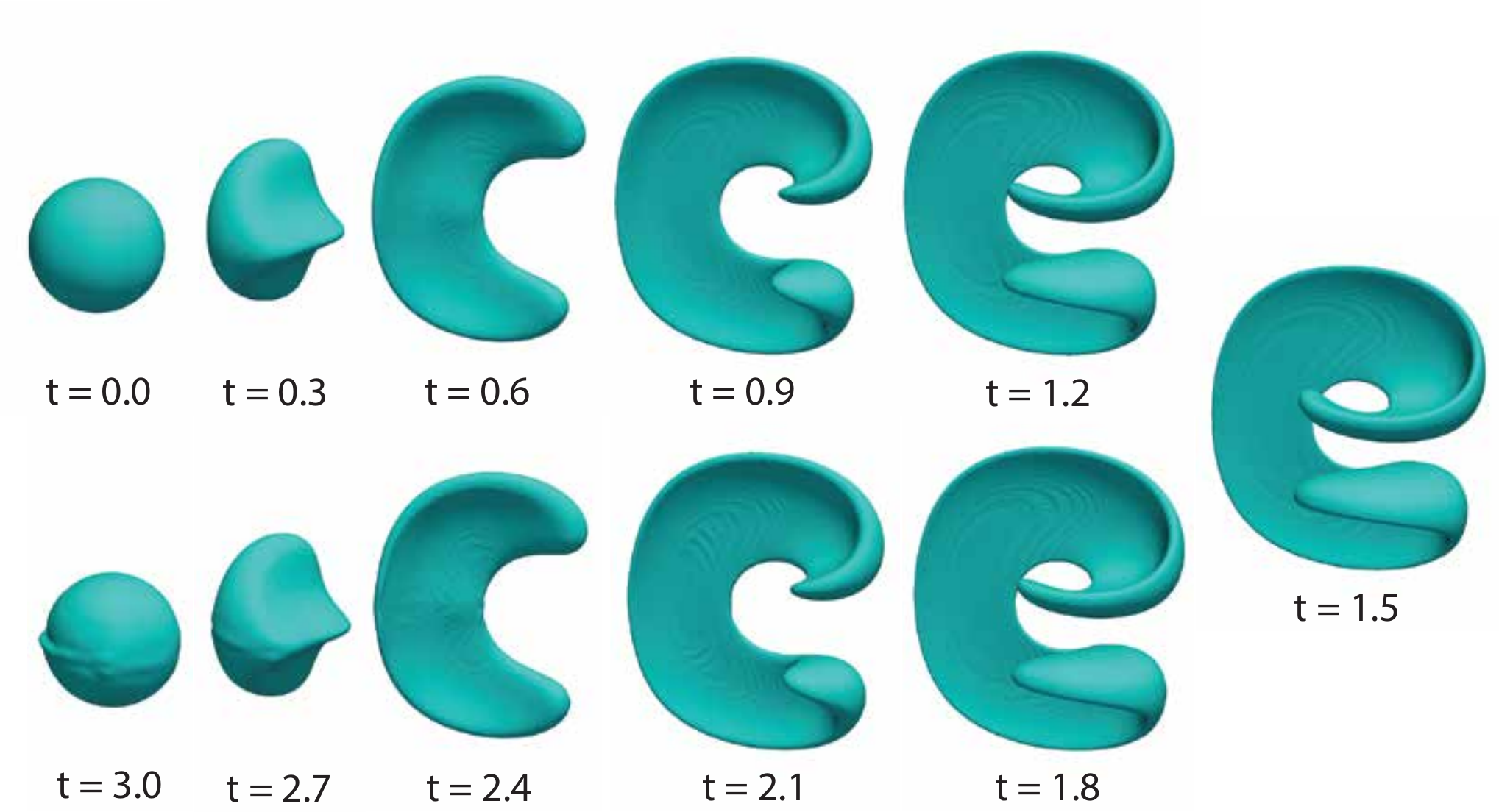}
\caption{Evolution of the three-dimensional sphere computed with the CLSVOF method on mesh size $1/256$}
\label{fig:SphereDeform}
\end{figure*}

Results were obtained using the CLSVOF algorithm on a $1/256$ mesh size grid. Figure \ref{fig:SphereDeform} shows the deformation experienced by the sphere over one time period. The deformed shape at $t=1.5$ corresponds to maximum stretching, while at $t=3.0$ the sphere has returned to its original position. The CLSVOF algorithm is able to resolve the thin stretched region at $t=1.5$. After the sphere has returned to its original position and shape, we observe a slight deviation from the initial spherical shape. The sphere develops a scar in the middle over one cycle of deformation. This deviation in shape is probably due to the accumulation of errors in the reconstruction of the interface at each time step. We compare the sphericity of a deformed sphere to quantify how accurately the sphere has retained its original shape. Sphericity is defined \citep{wadell1935volume} as:

\begin{equation}
    \label{sphericity}
    \mathrm{\Psi} = (\mathrm{\Gamma(t)})^{-1}\pi^{1/3}(6\mathrm{\Omega(t)})^{2/3}. 
\end{equation}

Here $\Gamma(t)$ is the surface area and $\Omega(t)$ is the volume of the sphere. Table \ref{tab:Sphericity} gives the value of sphericity for the deformation of the sphere. Sphericity, $\Psi = 1$ for perfect sphere. It is observed that initially, sphericity is almost equal to unity and it reaches a minimum as it stretches. Here sphericity at the final time step is very close to unity but still slightly lower than its initial value. This quantifies the deviation in the geometry of the sphere.\\

\begin{table*}
  \centering
  \caption{Sphericity of sphere undergoing deformation}
  \begin{tabular}{lccccc}
      \hline
      Time & 0.0 & 0.6 & 1.5 & 2.4 & 3.0 \\
      Sphericity & 0.9999 & 0.5036 & 0.2439 & 0.5025 & 0.9945\\
      \hline
  \end{tabular}
  \label{tab:Sphericity}
\end{table*}

\begin{figure*}
\centering
\includegraphics[width=0.6\textwidth,trim={0cm 0.4cm 0 0cm},clip] {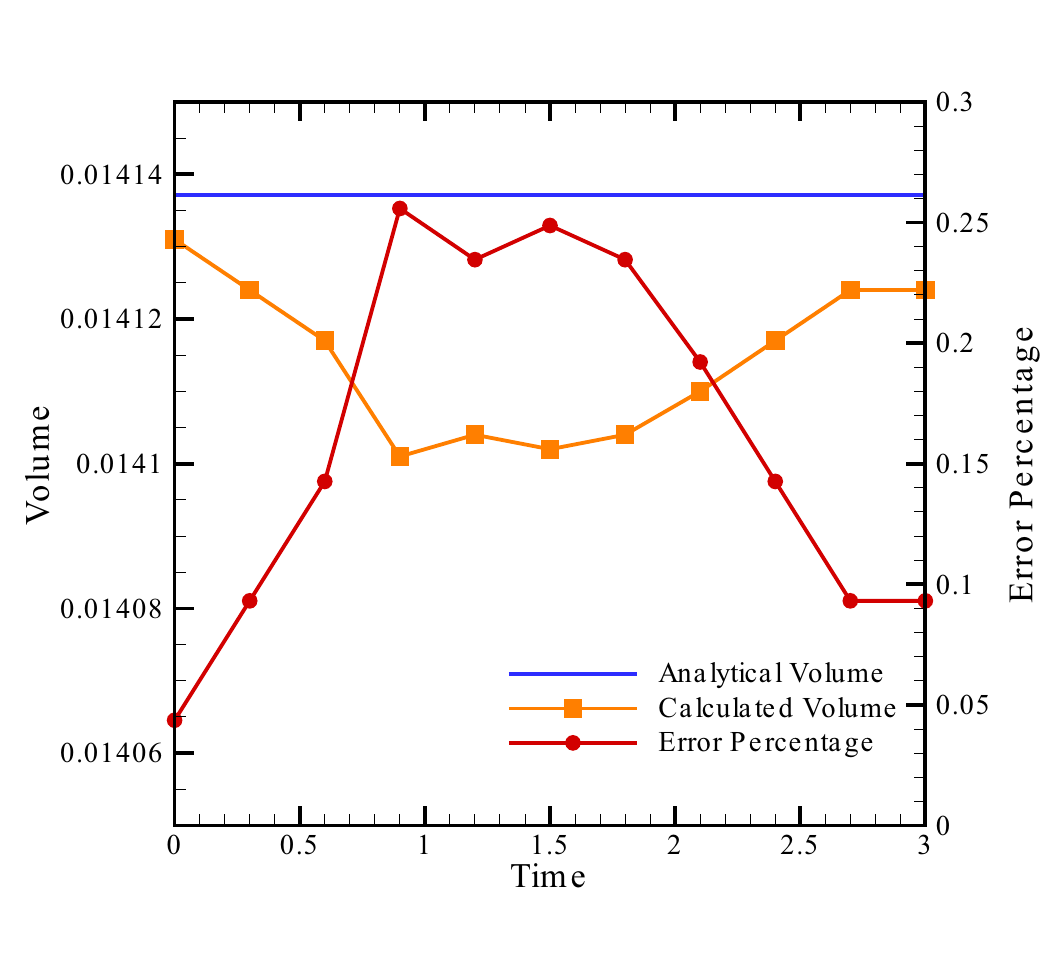}
\caption{Comparison between analytical volume and volume integral of CLSVOF results.}
\label{fig:mass_convergence}
\end{figure*}

Figure \ref{fig:mass_convergence} shows the volume convergence obtained from the CLSVOF algorithm over a cycle. As the sphere undergoes stretching, some volume is lost due to the numerical error in resolving the interface accurately. However as the sphere returns to its original shape, it recovers some of the lost volumes and gives relatively better volume conservation. The final error after one cycle is below $0.1\%$ which is in agreement with the results given by \cite{klitz_2015}.\\

\subsection{Rising Bubble in a Stratified Liquid Column}

A three-phase flow problem involving a bubble in a stratified liquid column with two liquids having different densities is used to further validate the numerical solver. An air bubble is placed inside the denser liquid and is allowed to rise gradually and interact with the interface. The physical properties of the fluids used are mentioned in table \ref{tab:RisingBubble}.\\

\begin{table*}
  \centering
  \caption{Physical properties of fluids for the rising bubble case}
  \begin{tabular}{p{0.20\linewidth}p{0.15\linewidth}p{0.15\linewidth}p{0.15\linewidth}p{0.15\linewidth}p{0.15\linewidth}}
  \hline
      \bf{Surface tension} & ($\mathrm{N}.\mathrm{m}^{-1}$) & \bf{Density} & ($\mbox{kg}.\mathrm{m}^{-3}$) & \bf{Viscosity} & ($\mbox{Pa}.\mathrm{s}$)\\
      $\sigma_{\mbox{gas - liquid}}$ & 0.07 & Bubble & 1 & Bubble & $10^{-4}$\\
      $\sigma_{\mbox{liquid - liquid}}$ & 0.05 & Heavy liquid & 1200 & Heavy Liquid & 0.15\\
      & & Light liquid & 1000 & Light Liquid & 0.1\\
      \hline
  \end{tabular}
  \label{tab:RisingBubble}
\end{table*}
For an air bubble rising in a stratified liquid column with two liquids, it can either get trapped at the interface of the liquids or penetrates the interface. There is also a possibility that the bubble entrains the heavy phase liquid if it does penetrate the interface. The condition for bubble penetrating the interface and the heavier liquid getting entrained is given by \cite{greene1988onset,greene1991bubble}.\\

\begin{figure*}
\centering
\includegraphics[width=1.0\textwidth,trim={0cm 0cm 0 0cm},clip] {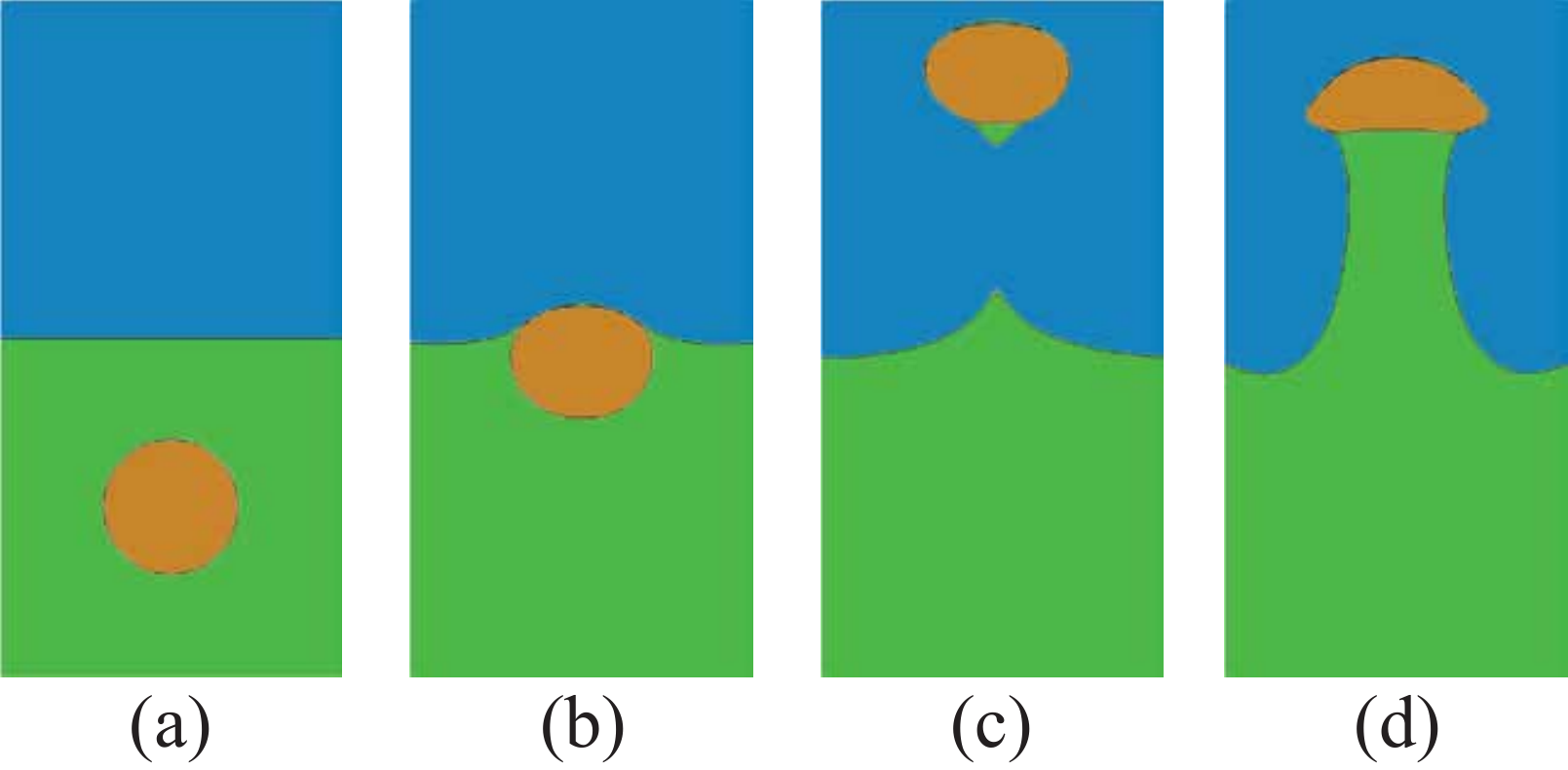}
  \caption{Simulation of rising bubble in a stratified liquid column. (\textit{a}) Initial state of bubble different bubble radii, (\textit{b}) Final state for bubble of radius $r=2mm$, (\textit{c}) Final state for bubble of radius $r=4mm$, (d) Final state for bubble of radius $r=8mm$.}
  \label{fig:RisingBubble}
\end{figure*}
Figure \ref{fig:RisingBubble}(a) shows the initial state of the bubble for the flow simulation performed using the CLSVOF algorithm. Initially, the lighter liquid stays above the heavier liquid and the bubble is placed some distance below the interface. Using the conditions for bubble penetration and fluid entrainment. \cite{greene1988onset,greene1991bubble, boyer2010cahn} gave different values of radii for different possible outcomes. Theoretically, the bubble would get trapped at the interface if the radius, $r<2.76mm$. Figure \ref{fig:RisingBubble}(b) shows the case for radius, $r=2.00mm$, and the bubble gets trapped at the interface. Figure \ref{fig:RisingBubble}(c) shows the case for a radius of bubble, $r=4.00mm$ which satisfies the condition for penetration and it can be seen that the bubble penetrates the interface with a very little entrainment of the heavier liquid. For the case with radius, $r=8mm$, figure \ref{fig:RisingBubble}(d), the bubble penetrates the interface while also entraining a large volume of the heavier liquid. The results obtained for the above three cases match with the results of \cite{boyer2010cahn} which was obtained using the Lattice Boltzmann method.\\

\section{Case setup}
\begin{figure}
\centering
\includegraphics[width=1.0\linewidth]{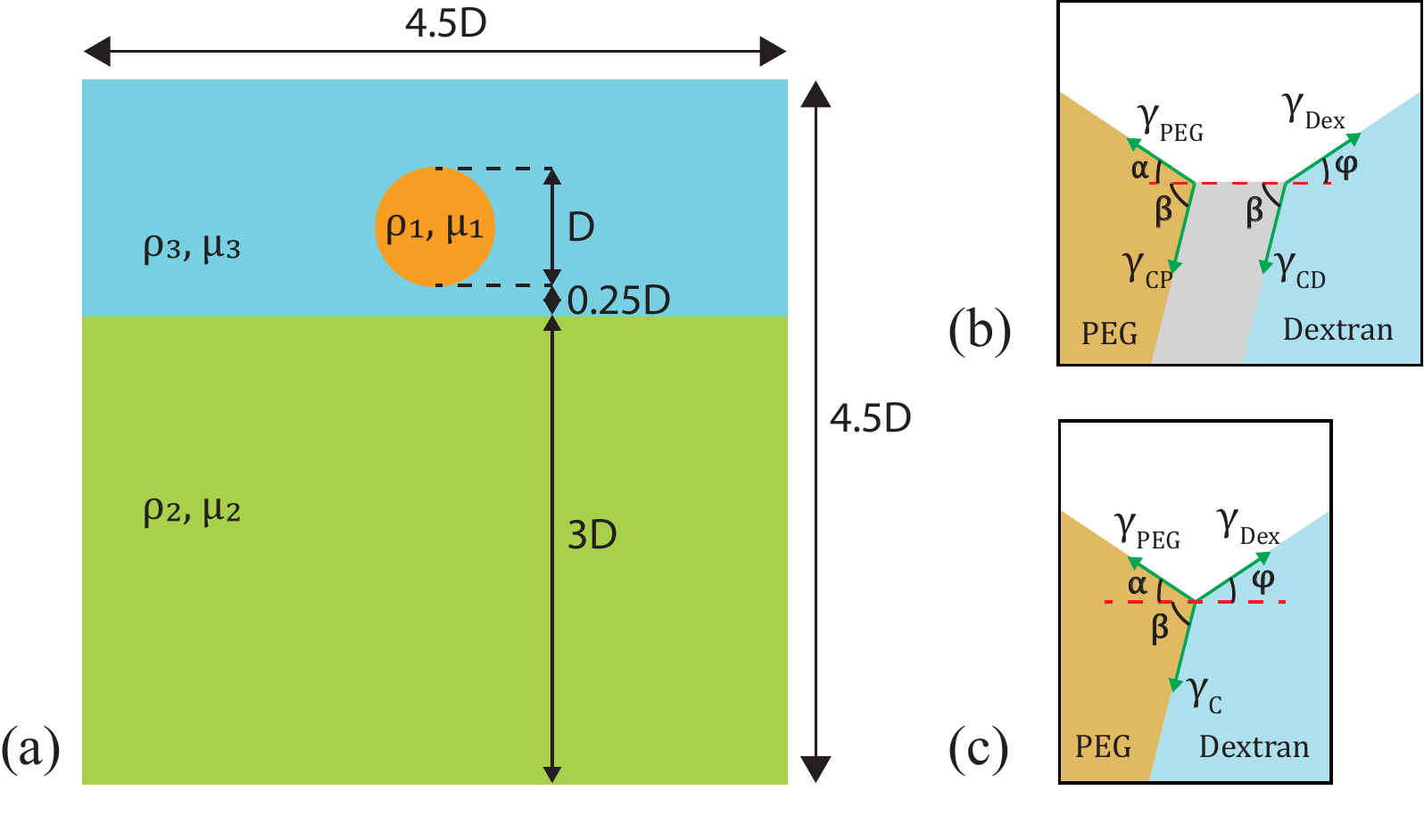}
\caption{(\textit{a}) Computational domain for the study of hanging droplets. Coacervate model and contact angles at the triple phase contact line,  (\textit{b}) interfacial tension at the coacervate with two surfaces and finite coacervate thickness, and (\textit{c}) representation of the coacervate with a single surface.}
\label{fig:dom}
\end{figure}

Figure \ref{fig:dom} (a) shows the computational domain used in the present simulations. A spherical droplet of diameter D is placed slightly above the centre of the cubical computational domain of side length $4.5$D. The depth of the pool is taken as $3$D in order to ensure that the droplet is sufficiently far away from the computational boundaries. In the experiments \cite{xie2020}, the droplets were released from varying heights but in order to minimize the computational domain size, the droplets are released from a fixed height of $0.25$D but with different initial velocity. The impact of droplet is considered for very low Reynolds numbers and Weber numbers. Therefore, splashing and jets are not expected during this impact. All the simulations are performed on a grid size of $1/128$ in all the three directions.\\

The droplet containing the dextran solution is taken as phase 1, the pool of PEG solution is taken as phase 2, and the air is taken as phase 3 for the three-phase flow solver. Figure \ref{fig:dom} (b) shows the coacervate layer between the two immiscible solutions of dextran and PEG. $\gamma_{Dex}$ and $\gamma_{PEG}$ are the surface tension values for the dextran and the PEG solution. The interfacial tension at the coacervate-dextran and coacervate-PEG interface is given by $\gamma_{CD}$ and $\gamma_{CP}$. Generally, the coacervate thickness is assumed to be very small and for the simplicity of modeling, it is taken as a single surface. The two interfacial tension at the coacervate are combined to give a single interfacial tension at the dextran-PEG interface, ($\gamma_C=\gamma_{CD}+\gamma_{CP}$), as shown in figure \ref{fig:dom} (c). The physical properties of the fluids used in the numerical simulation are given in table \ref{tab:HangingDrop}.\\

\begin{table*}
  \centering
   \caption{Physical properties of fluids for the hanging droplet case}
  \begin{tabular}{p{0.35\linewidth}p{0.25\linewidth}p{0.35\linewidth}}
  \hline
      \bf{Surface tension} ($\mathrm{N}.\mathrm{m}^{-1}$) & \bf{Density} ($\mbox{kg}.\mathrm{m}^{-3}$) & \bf{Viscosity} ($\mbox{Pa}.\mathrm{s}$)\\
      $\sigma_{12}=\gamma_{C}=0.02$ & $\rho_1=\rho_{Dex}=1055$ & $\mu_1=\mu_{Dex}=6.0\times10^{-2}$\\
      $\sigma_{13}=\gamma_{Dex}=0.0356$ & $\rho_2=\rho_{PEG}=1014$ & $\mu_2=\mu_{PEG}=7.05\times10^{-3}$\\
      $\sigma_{23}=\gamma_{PEG}=0.0256$& $\rho_3=\rho_{Air}=1.3$ & $\mu_3=\mu_{Air}=1.6\times10^{-5}$\\
      \hline
  \end{tabular}
  \label{tab:HangingDrop}
\end{table*}
\begin{figure}
\centering
\includegraphics[width=1.0\linewidth]{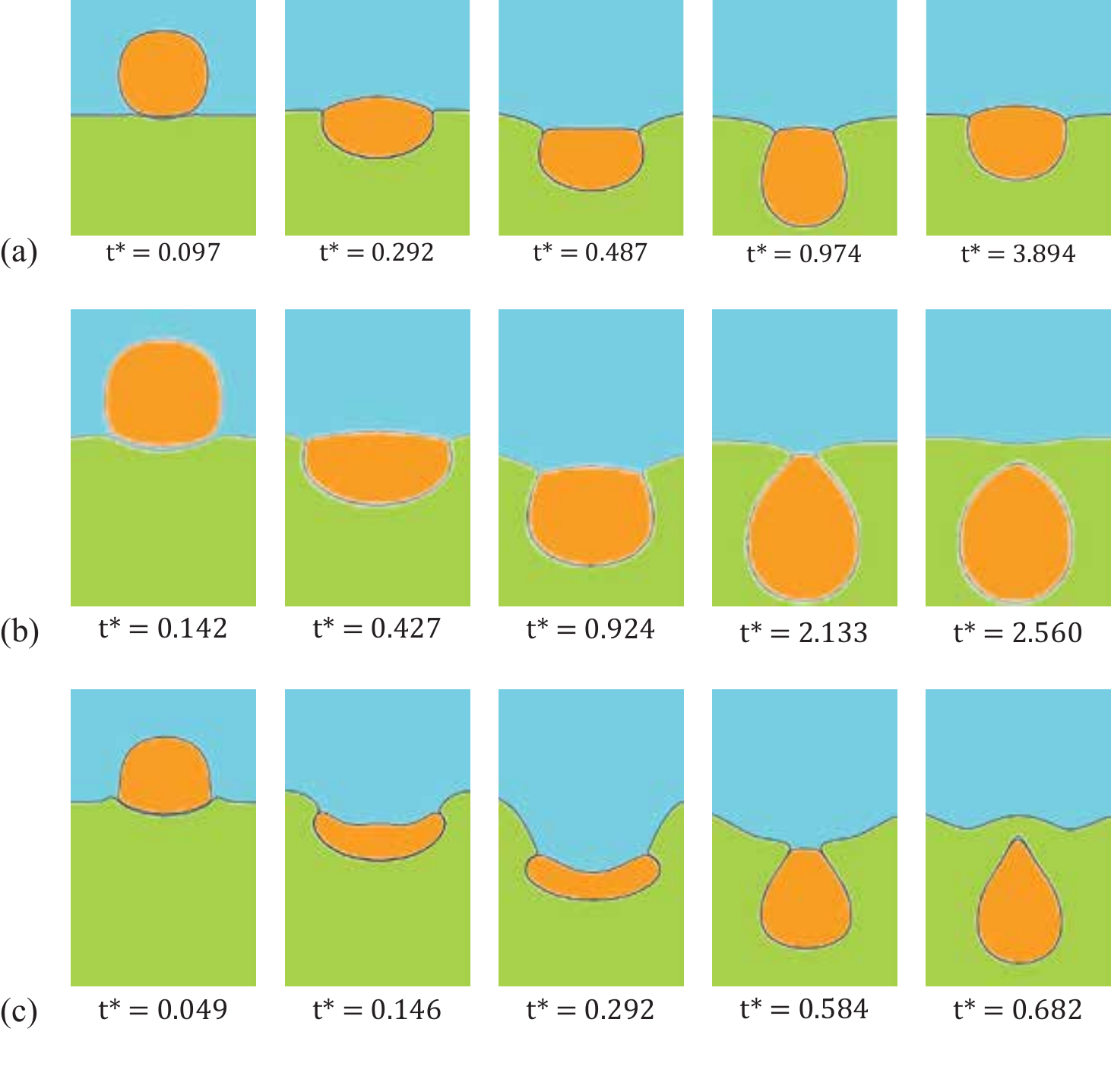}
\caption{Evolution of (\textit{a}) hanging droplet for  2mm diameter released from 4.98 mm height, (\textit{b}) intermediate droplet for 0.6 mm diameter released from 61.67 mm height, and (\textit{c}) wrapping droplet for 2 mm diameter released from 50.97 mm height.}
\label{fig:HD591}
\end{figure}

\section{Results}

\subsection{Hanging, Intermediate and Wrapping Droplets}
\begin{figure}
\centering
\includegraphics[width=1.0\linewidth]{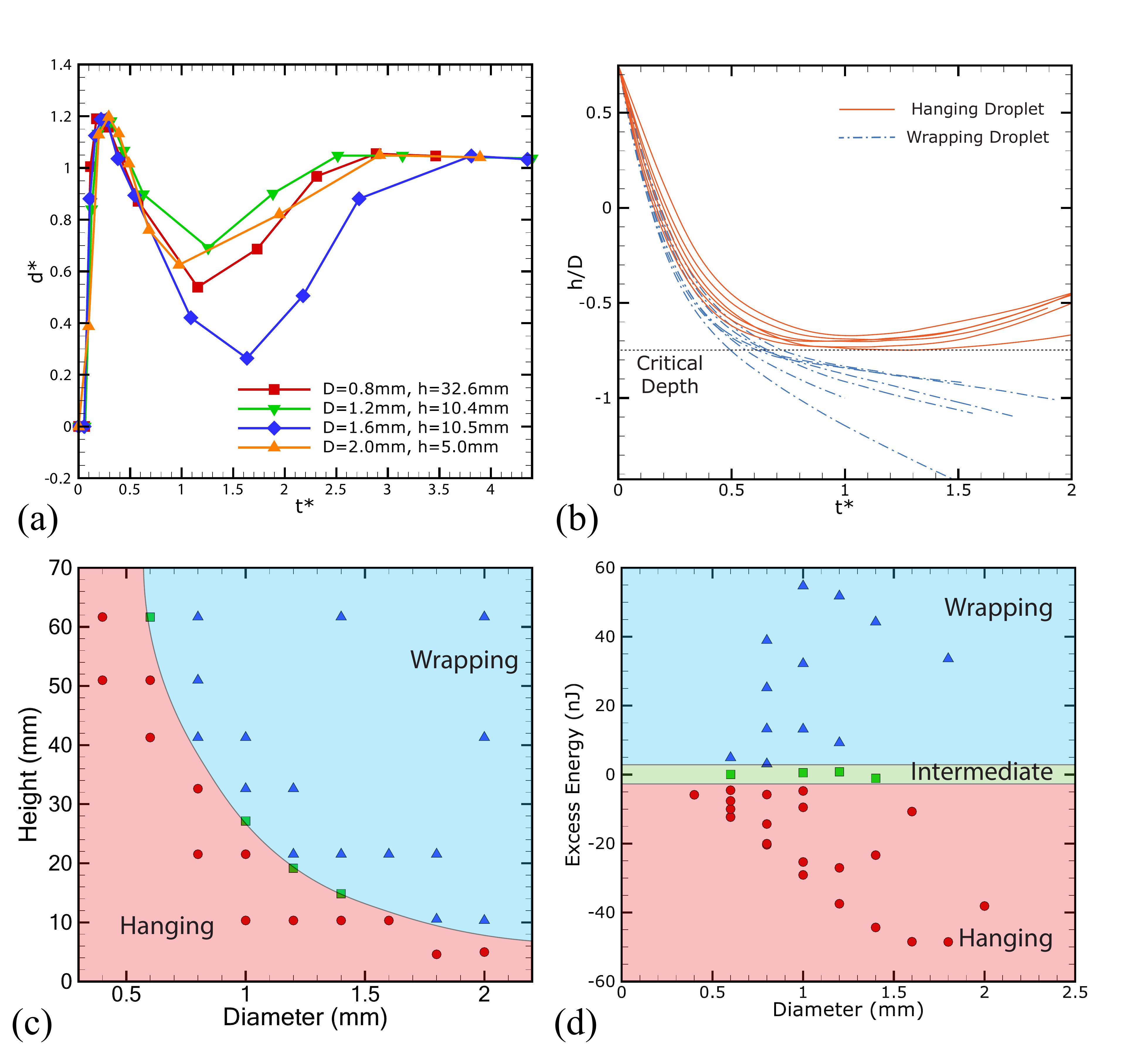}
\caption{(\textit{a}) Variation of non-dimensional TPCL diameter with respect to the non-dimensional time for different hanging droplet cases. Here $d^*$ is defined as $d^*=d/D$ where $d$ is the TPCL diameter. (\textit{b}) Variation of non-dimensional height of the droplet for different cases of hanging and wrapping droplets. State diagram for hanging and wrapping droplets as a function of (\textit{c}) height and diameter, (\textit{d})  excess energy and diameter.}
\label{fig:heights}
\end{figure}

We perform three-dimensional numerical simulations for the above-mentioned configuration and found that the heavier droplet hangs from the lighter liquid interface for certain diameters and impact velocity of the droplet. Figure \ref{fig:HD591} (a) shows the evolution of a hanging droplet (Movie S1) upon impact of a drop of diameter, $D=2$ mm, released from a height of $h = 4.98$ mm (impact velocity of $\sqrt{2gh} = 0.313$ m/s). Here, $t^*=t/\tau_c$, capillary time $\tau_c$ is defined as $\tau_c=\sqrt{\rho_1 D^3/\sigma_{12}}$. As the droplet makes a transition from hanging to sinking, an intermediate case (Movie S4) is also observed as shown in figure \ref{fig:HD591} (b) for a droplet of $0.6$ mm diameter released from a height of $h = 61.67$ mm (impact velocity of $\sqrt{2gh} = 1.1$ m/s). Figure \ref{fig:HD591} (c) shows a case for a $2$ mm  diameter released from a height of $h = 50.97$ mm (impact velocity of  $\sqrt{2gh} = 1$ m/s) in which the droplet sinks into the pool upon impact (Movie S7).  It is observed that the droplet begins to slow down even before it makes contact with the pool. As the drop moves closer to the pool, a thin film of air separates the droplet \citep{duchemin2020dimple} from the pool. This acts as a cushion and is responsible for the decrease in the impact velocity of the droplet. As the droplet makes contact with the pool, the TPCL diameter expands rapidly. After the impact, the droplet drastically loses its kinetic energy by displacing a portion of the pool towards the pool surface. This creates a crater in the pool shrinking the TPCL diameter. The droplet sits in this crater and hangs from the surface. After the droplet loses all its kinetic energy, it starts moving upwards and keeps oscillating with very small amplitude until it reaches an equilibrium height. The TPCL diameter again increases during this process. The evolution of the non-dimensional TPCL diameter with respect to the non-dimensional time for different hanging droplet cases is shown in \ref{fig:heights}(a). A capillary wave \citep{che2018impact} is formed upon the impact of the droplet. It is also observed that the equilibrium height and the shape of the hanging droplet are independent of the release height of the droplet as long as it hangs from the surface. This independence of the final shape of the droplet on the impact velocity or release height differs from the results presented by \cite{xie2020} owing to the representation of the coacervate with a single surface.\\

Figure \ref{fig:heights} (b) shows the variation of the height of the center of mass of the droplet from the pool surface for various cases. It can be observed that there exists a critical depth upon crossing which the droplet gets wrapped. Droplets that do not cross this critical depth tend to hang from the pool surface. The critical depth is found to be $0.74$ times the diameter of the droplet. It is to note that the critical depth is greater than half the diameter of the droplet, i.e. for a brief moment the droplet goes completely below the pool surface displacing the pool fluid. Since the computational domain is taken as a closed container such that no fluid exits the domain, the displaced fluid increases the pool height. Increased pool height results in additional pressure head which pushes the droplet upwards. However, if the pool height increases significantly then it covers the top surface of the droplet and wraps it completely. The droplet sinks when it gets wrapped by the pool fluid.\\

Simulations for different droplet diameters and release heights are performed (see Movies S1-S9). It is observed that the tendency of a droplet to hang from the pool surface increases as the droplet radius or the release height is reduced. A droplet of a diameter of $2$ mm released from a height of $10$ mm gets wrapped and sinks into the pool. In contrast, a droplet of a diameter of $0.4$ mm released from a height of even $60$ mm hangs from the pool surface. Figure \ref{fig:heights} (c) shows both the hanging and wrapping state as a function of the droplet diameter and the release height. A non-linear curve divides both the states. There are also a few cases that lie very close to the curve dividing the two states. These are the cases where droplet upon impact with the pool briefly gets stuck at the pool surface and slowly moves downwards eventually sinking into the pool. The cases close to the curve dividing the two states are the intermediate cases.\\

\subsection{Energy Balance For Hanging Droplets}

Empirical energy calculations are performed to justify hanging and wrapping states for different cases. It is assumed that the droplet is released from height $h$ and the entire potential energy is converted into kinetic energy at the time of impact, 
\begin{equation}
    E_{KE}=\frac{\pi}{6}D^3\rho_1 gh.
\end{equation}
This is the entire energy available with the droplet which is used to overcome different forms of energy requirements. Three different forms of energy losses are considered here for energy balance. Firstly, a part of the available energy is spent to displace the pool fluid to the pool surface to create a crater for the droplet. It is observed from figure \ref{fig:heights} (b) that if a droplet is getting wrapped, it needs to attain a critical depth. From this, the displaced volume $\Delta \mathcal{V}$ is approximated as
\begin{equation}
    \Delta \mathcal{V} \approx \frac{\pi}{4}D^2\times 0.74D + \frac{\pi}{12}D^3.
\end{equation}
Here it is assumed that the crater is cylindrical shaped with a hemisphere at one of its end. The center of mass of the crater is given as 
\begin{equation}
    z_c = \frac{\frac{\pi}{4}D^2\times 0.74D \times \frac{0.74}{2}D + \frac{\pi}{12}D^3\times \left(0.74+\frac{3}{16}\right)D}{\frac{\pi}{4}D^2\times 0.74D + \frac{\pi}{12}D^3} 
\end{equation}
%Therefore the center of mass of the crater, $z_c = 0.5431D$. 
This $z_c = 0.5431D$ is the height by which the crater needs to be lifted and the energy required for this is calculated as potential energy loss.
Taking a correction factor $k_{PE}$, the potential energy loss is evaluated as 
\begin{equation}
    E_{PE} \approx k_{PE} \times \rho_2 \Delta \mathcal{V}gz_c = k_{PE} \times 4.4916\rho_2D^4.
\end{equation}
The correction factor for the potential energy loss is taken as unity. By taking the shape of the crater as defined above, change in surface area can be evaluated for different surfaces. Taking the product of these surface changes with their respective surface tension values gives an estimate of the energy required for the destruction and creation of new surfaces. 
\begin{equation}
    E_{S} \approx \left( 0.74\pi D^2\sigma_{23} + \frac{\pi}{2}D^2\sigma_{12}\right)-\left(\frac{\pi}{2}D^2\sigma_{13} + \frac{\pi}{4}D^2\sigma_{23} \right)
\end{equation}
Using the values of $\sigma_{12}, \sigma_{23}$ and $\sigma_{13}$ this energy is further approximated as 
\begin{equation}
    E_{S}  \approx k_S \times 0.005\pi D^2
\end{equation}
%Here we take $k_S=1.4$. {\color{blue} Because 
The actual crater is not exactly cylindrical, but rather has curved edges and capillary waves. Therefore, the actual surface generated should be bigger than estimated. As a result, the adjustment factor $k_S$ should be greater than $1$.
The PEG solution in the pool is a highly viscous fluid and hence large viscous losses are also expected due to the motion of droplet into the pool. Since the force experienced by a droplet when moving through another fluid is not known exactly, following assumptions are made to approximate this energy loss: (a) the droplet is assumed to be a rigid sphere moving through the pool, (b) flow speed past the droplet is taken as constant. Under these assumptions the drag force experienced by the droplet is given by
\begin{equation}
    F_d = \frac{1}{2}C_d\rho_2 V^2 \times A,
\end{equation}
where $A=\pi D^2/4$ is the frontal area. The distance travelled by the droplet $z_d$ is the sum of the height of the center of mass of the drop at time of impact above the interface ($0.5D$) and the critical depth ($0.74D$). We also have to include a correction factor $k_V$ to accommodate the aforementioned assumptions. This gives the viscous losses as
%\begin{equation}
%    E_V = k_V \times \frac{1}{2}C_d\rho_2 V^2 \times 1.24\pi D^3
%\end{equation}
\begin{equation}
    E_V  \approx k_V \times \frac{1}{2}C_d\rho_2 V^2 A z_d = 0.155 k_V C_d\rho_2 V^2\pi D^3
\end{equation}
%We take $k_V=0.88$. 
When the droplet descends from the interface, its velocity decreases, and its shape changes, resulting in a decrease in the drag coefficient. Therefore, the overall viscous losses will be lower than the estimated value and should also be accounted for by the correction factor. Here, $C_d$ is the drag coefficient for flow past a sphere and is dependent on the droplet diameter D, impact velocity V, the density $\rho_2$, and viscosity $\mu_2$ of the pool. The Reynolds number, $Re = \frac{\rho_2 V D}{\mu_2}$ is $1<Re<1000$. Hence, the drag coefficient is defined using the relation given by Schiller and Naumann \citep{flemmer1986drag}. Considering the above-mentioned four energies, it is determined whether a droplet, if it crosses the critical depth, still has additional energy to move further downwards. Excess energy is calculated as
\begin{equation}
    E_{\mbox{excess}} = E_{KE}-E_{PE}-E_S-E_V
\end{equation}
The values of the correction factors $k_S$ and $k_V$ are tuned such that the available data set for the final state of droplet impact gives a distinct distribution in terms of the excess energy. Figure \ref{fig:heights} (d) shows the state diagram for hanging and wrapping droplets as a function of excess energy and diameter of the droplet. It can be seen that the droplets with sufficient energy to spend on different losses tend to get wrapped and sink into the pool whereas the droplets which have less energy, to begin with, such that they have negative excess energy, tend to hang from the pool surface. There are also intermediate cases where the available energy is nearly equal to the energy required and thus has close to zero excess energy. These droplets initially lose their entire kinetic energy upon impact and then gradually sink into the pool. It is observed that the majority portion of the available energy is spent to overcome the viscous loss and the remaining energy is spent for surface energy. A very small part of the available energy is spent on the potential energy loss. Thus a larger droplet with higher initial energy can still hang from the surface if either the viscosity of the pool fluid is increased or the interfacial tension value used for the coacervate is increased. It is to note here that the excess energy is just a function of D and h, and it converts the non-linear distribution of hanging and wrapping droplets in figure \ref{fig:heights} (c) into a linear distribution in figure \ref{fig:heights} (d).\\

\subsection{Force balance for hanging droplets}
The viscous force has a significant impact on the droplet's rate of descent; nevertheless, after it has reached equilibrium, it is the surface tension force and the buoyant forces that are responsible for maintaining the droplet's attachment to the surface by balancing its weight. We present the calculations for the force balance based on an analytical approach and the outcomes of the numerical simulations. The weight of the droplet is calculated as
\begin{equation}
    F_{w} = \mathcal{V}_1 \times \rho_1 \times g.
\end{equation}
Here $\mathcal{V}_1$ is the total volume of the droplet. The buoyant force is defined as 
\begin{equation}
    F_{b} = \mathcal{V}_2 \times \rho_2 \times g.
\end{equation}
$\mathcal{V}_2$ here is the volume of pool fluid displaced by the droplet below TPCL. It is worth noting that the droplet at equilibrium is not completely immersed beneath the pool's surface. A little portion of the droplet remains above the TPCL line. There are also pockets of air bubbles trapped between the droplet and pool interfaces. The volume of these air bubbles is also included in $\mathcal{V}_2$ to calculate the buoyant forces. Now consider a system with a droplet including the droplet-air interface and the droplet-pool interface. The forces acting on the system in the vertical direction are only the gravitational force, the buoyant force, and the surface tension due to the air-drop interface and drop-pool interface. Since the hanging droplet system consists of two different surfaces wrapped around a common ring, i.e. the TPCL, the surface tension forces due to each of the two interfaces can be evaluated by taking the product of pressure jump across the interface and the projected area at their boundary. Pressure jump at the interface can be calculated using the Young's Laplace equation,
\begin{equation}
    \Delta p_{ij} = \frac{2\sigma_{ij}}{R_{ij}},
    \label{eq:Young_Laplace}
\end{equation}
here $R_{ij}$ is the radius of curvature of the interface between the phases i and j. Therefore, the vertical force on the hanging droplet owing to the surface tension computed from the values of $R_{12}$ and $R_{13}$ obtained from the simulations is given by,
\begin{equation}
    F_{\sigma} = \left(\frac{2\sigma_{12}}{R_{12}}-\frac{2\sigma_{13}}{R_{13}}\right) \times \frac{\pi d^2}{4}.
    \label{eq:ST_Force}
\end{equation}
Based on the values of $\mathcal{V}_1$ and $\mathcal{V}_2$ obtained from our simulations, in order to satisfy the force balance on the hanging droplet in the vertical direction, the ideal value of vertical component of surface tension force should be
\begin{equation}
    F_{\sigma0} = F_w-F_b.
    \label{eq:ForceBal}
\end{equation}
$F_{\sigma}$ computed from the simulations are in fact very close to $F_{\sigma0}$ as demonstrated in table \ref{tab:Force} for four cases of hanging droplets signifying the dynamical balance. \\

\begin{table*}
  \centering
  \caption{Calculation for surface tension force balance for the hanging droplet.}
  \begin{tabular}{p{0.35\linewidth}p{0.15\linewidth}p{0.15\linewidth}p{0.15\linewidth}p{0.15\linewidth}}
\hline
       Droplet diameter, $D$ ($m$) & $2.00\times 10^{-3}$ & $1.60\times 10^{-3}$ & $1.20\times 10^{-3}$ & $0.80\times 10^{-3}$\\
       Droplet release height, $h$ ($m$) & $5.00\times 10^{-3}$ & $10.50\times 10^{-3}$ & $10.40\times 10^{-3}$ & $32.60\times 10^{-3}$\\
       Total volume of the droplet, $\mathcal{V}_1$ $(m^3)$ & $4.19\times 10^{-9}$ & $2.14\times 10^{-9}$ & $9.05\times 10^{-10}$ & $2.68\times 10^{-10}$\\
       Displaced volume from pool, $\mathcal{V}_2$ $(m^3)$ & $3.59\times 10^{-9}$ & $1.84\times 10^{-9}$ & $7.64\times 10^{-10}$ & $2.30\times 10^{-10}$\\
       Weight of droplet, $F_w$ $(N)$ & $4.30\times 10^{-5}$ & $2.26\times 10^{-5}$ & $9.49\times 10^{-6}$ & $2.81\times 10^{-6}$\\
       Buoyant force, $F_b$ $(N)$ & $3.58\times 10^{-5}$ & $1.83\times 10^{-5}$ & $7.60\times 10^{-6}$ & $2.28\times 10^{-6}$\\
       Surface tension force, $F_{\sigma0}$ $(N)$ & $7.60\times 10^{-6}$ & $3.85\times 10^{-6}$ & $1.77\times 10^{-6}$ & $4.91\times 10^{-7}$\\
       TPCL diameter, $d$ $(m)$ & $2.08\times 10^{-3}$ & $1.65\times 10^{-3}$ & $1.24\times 10^{-3}$ & $0.84\times 10^{-3}$\\
       \bf{Radius of Curvature, $(mm)$} & & & &\\
       $\quad R_{12}$ & $1.076$ & $0.868$ & $0.651$ & $0.444$\\
       $\quad R_{13}$ & $2.140$ & $1.658$ & $1.210$ & $0.806$\\
       \bf{Pressure Jump, $(Pa)$} & & & &\\
       
       $\quad\Delta p_{12}$ & $37.186$ & $46.075$ & $61.467$ & $90.020$\\
       
       $\quad\Delta p_{13}$ & $34.932$ & $44.244$ & $59.970$ & $89.128$\\
       
       \bf{Surface tension force $F_{\sigma}$ $(N)$} & $7.66\times 10^{-6}$ & $3.91\times 10^{-6}$ & $1.81\times 10^{-6}$ & $4.95\times 10^{-7}$\\
       
       \hline
  \end{tabular}
  \label{tab:Force}
\end{table*}

\begin{figure*}
\centering
\includegraphics[width=1.\textwidth]{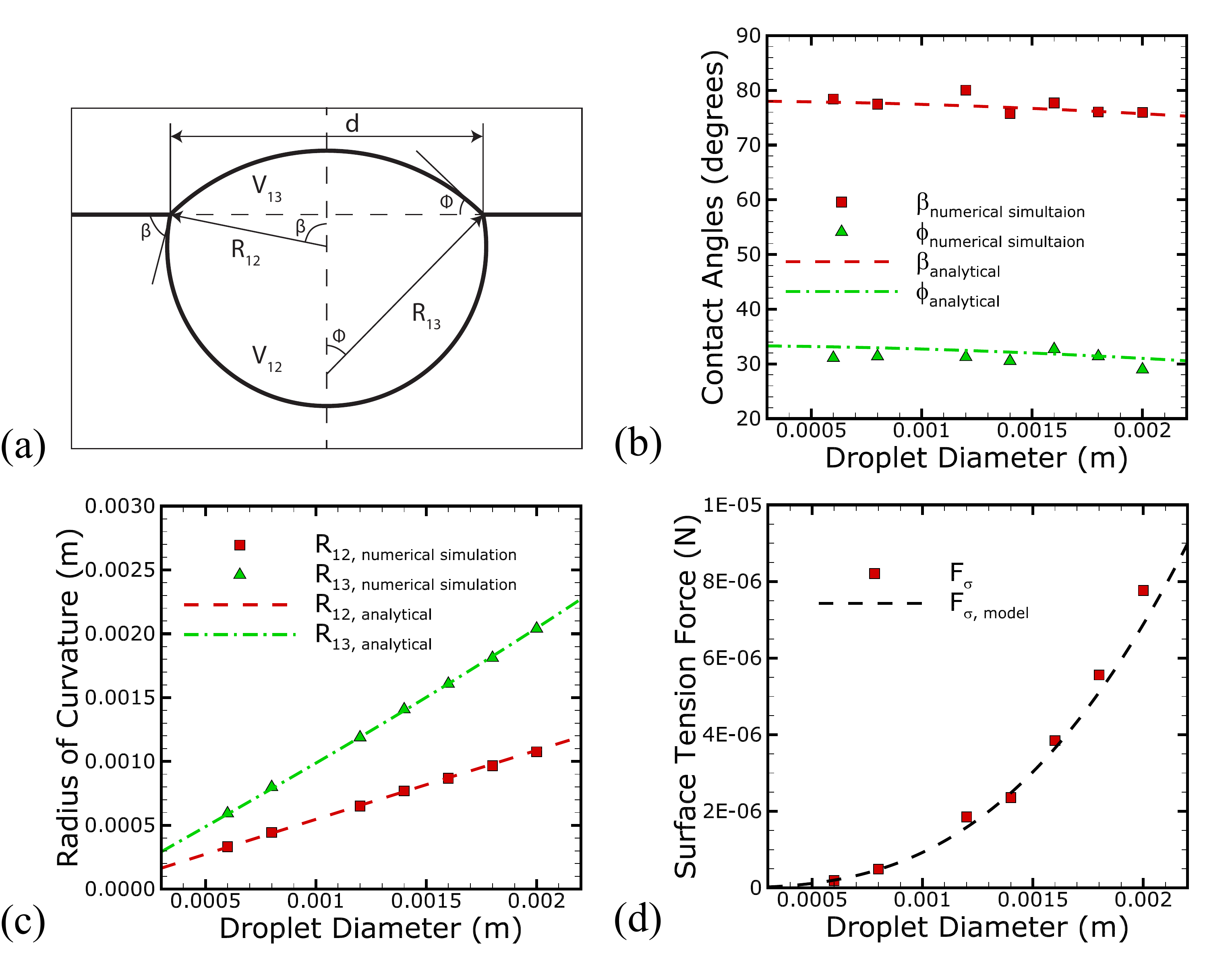}
\caption{(\textit{a}) Shape of droplet at equilibrium, (\textit{b}) contact angles $\beta$ and $\phi$ vs diameter of the droplet at equilibrium, (\textit{c}) radius of curvature for the interface vs diameter of the droplet at equilibrium, and (\textit{d}) surface tension force on the droplet vs diameter of the droplet at equilibrium.}
\label{fig:dropmodel}
\end{figure*}

Our simulations show that the shape of a hanging droplet at equilibrium resembles a combination of two spherical caps with varying radii. Hence, we model the droplet with two spherical sections of radii $R_{12}$ and $R_{13}$ respectively as shown in figure \ref{fig:dropmodel} (a). Considering the two interfaces as part of purely spherical sections, the radius of curvature and the TPCL diameter can be related as 
\begin{equation}
    d = 2R_{12}\sin\beta = 2R_{13}\sin\phi.
    \label{eq:TPCL_curvature}
\end{equation}
Using equation \ref{eq:TPCL_curvature}, volume of the upper and lower spherical sections, $V_{13}$ and $V_{12}$ can be evaluated as follows:
\begin{equation}
    V_{13} = \frac{\pi}{24\sin^3\phi}d^3(2+\cos\phi)(1-\cos\phi)^2,
    \label{eq:v13}
\end{equation}
\begin{equation}
    V_{12} = \frac{\pi}{24\sin^3\beta}d^3(2-\cos\beta)(1+\cos\beta)^2.
    \label{eq:v12}
\end{equation}
The droplet and the pool fluids are taken as immiscible because there is no chemical reaction taking place at the interface. Therefore, the volume of the droplet must be conserved  
\begin{equation}
    V_{12}+V_{13} = \frac{\pi}{6}D^3.
    \label{eq:vol_consrv}
\end{equation}
Again using \ref{eq:TPCL_curvature} in \ref{eq:ST_Force}, the vertical surface tension force on the droplet is calculated analytically as 
\begin{equation}
    F_{\sigma,Model} = \pi d (\sigma_{12}\sin\beta-\sigma_{13}\sin\phi).
    \label{eq:SFT_model}
\end{equation}
However, considering the vertical force balance on the droplet, i.e.\ using \ref{eq:ForceBal}, \ref{eq:v12} and \ref{eq:v13}, we get
\begin{equation}
    F_{\sigma,Model} = \left((\rho_1-\rho_2)V_{12}+(\rho_1-\rho_3)V_{13}\right)g.
    \label{eq:ForceBal_model}
\end{equation}
Apart from the force balance on the droplet, the interfaces between the three phases, air, droplet and pool are also considered to be massless. Therefore, at the junction of the three phases, i.e.\ at the TPCL the vertical and the horizontal surface tension forces due to the three interfaces must balance each other. Therefore, 
\begin{equation}
    \sigma_{23}\sin\alpha-\sigma_{12}\sin\beta+\sigma_{13}\sin\phi = 0,
    \label{eq:Ver_ForceBal}
\end{equation}
\begin{equation}
    \sigma_{23}\cos\alpha+\sigma_{12}\cos\beta-\sigma_{13}\cos\phi = 0.
    \label{eq:Hor_ForceBal}
\end{equation}
Eliminating $\alpha$ (see figure \ref{fig:dom}(b)) from \ref{eq:Ver_ForceBal} and \ref{eq:Hor_ForceBal}, we get
\begin{equation}
    (\sigma_{13}\sin\phi-\sigma_{12}\sin\beta)^2 + (\sigma_{13}\cos\phi-\sigma_{12}\cos\beta)^2 = \sigma^2_{23}
    \label{eq:beta_phi}
\end{equation}
Equations \ref{eq:vol_consrv}, \ref{eq:SFT_model}, \ref{eq:ForceBal_model} and \ref{eq:beta_phi} can be solved simultaneously to obtain the values for $\beta,\phi,d$ and $F_{\sigma,Model}$. Figures \ref{fig:dropmodel} (b), (c) and (d) show an excellent match of the contact angles ($\beta$ and $\phi$), radii of curvature for the two interfaces and the surface tension forces respectively between our simulations and the analytical solution obtained by solving the force balance equations for the hanging drops.\\

\section{Conclusions}
In this work, the impact of a droplet into a pool of immiscible liquid is investigated using three-dimensional three-phase flow simulations. The results from the numerical simulations suggest that the droplet upon impact can either hang from the liquid surface or get wrapped into the pool and sink eventually. In some rare cases, the droplet even gets stuck at the interface and gradually sinks into the pool. All three cases obtained from the numerical results are shown to happen in experiments \cite{xie2020} as well. Further, a parametric study of the droplet impact is done to understand the effect of droplet diameter and release height on the final state of the droplet. It is observed that a non-linear curve in terms of droplet diameter and release height separates the hanging and wrapping state. As the droplet diameter or the release height is increased, the droplets move from the hanging state to the wrapping state. It is observed that the shape of the droplet at equilibrium does not vary with release height for hanging droplets. A hanging droplet of a given diameter tends to have a unique final state. This behavior of the hanging droplets is different from the observation of \cite{xie2020}. The simplicity of the model used for the coacervate is the probable reason for this divergence from the experimental results. This suggests that the coacervate needs more sophisticated modeling for its physical properties even if it is considered to have no mass. \\

Further, an approximate energy balance is presented for the droplet impact. It is shown that the major portion of the kinetic energy available with the droplet is dissipated as viscous losses. The remaining energy is converted to the surface and potential energy owing to the formation of the crater. The energy balance is then used to determine whether a given heavier droplet will float or sink in the pool of the lighter liquid. Additionally, we solve the force balance equations of a hanging drop analytically at the equilibrium position. The values of the contact angle, the radii of curvature, and the force due to surface tension at the TPCL obtained from this dynamical balance show an excellent match with the simulations. \\

A natural extension of this work will be to perform a parametric study to understand the effect of other fluid parameters such as the viscosity and the surface tension values. Furthermore larger sized droplets with much higher energy can be simulated to potentially get some new states like droplets breaking off of the surface leaving a secondary droplet at the surface.\\

\backsection[Supplementary data]{\label{SupMov} Supplementary movies of hanging, intermediate and wrapping droplets are available at \\https://doi.org/**.****/jfm.***...}

\backsection[Acknowledgements]{We gratefully acknowledge the support of the Science and Engineering Research Board, Government of India grant no. SERB/ME/2020318. We also want to thank the Office of Research and Development, Indian Institute of Technology Kanpur for the financial support through grant no. IITK/ME/2019194. The support and the resources provided by PARAM Sanganak under the National Supercomputing Mission, Government of India at the Indian Institute of Technology, Kanpur are gratefully acknowledged.}

%\backsection[Funding]{This research received no specific grant from any funding agency, commercial or not-for-profit sectors.}

\backsection[Declaration of interests]{ The authors report no conflict of interest.}

%\backsection[Data availability statement]{The data that support the findings of this study are openly available in [repository name] at http://doi.org/[doi], reference number [reference number].}

%\backsection[Author ORCID]{Souvik Naskar, https://orcid.org/0000-0003-0445-8417; Anikesh Pal, https://orcid.org/****-****-****-****}

\backsection[Author contributions]{P.S. and A.P. designed research; P.S. and N.S. developed the numerical solver, P.S. and A.P. performed research; P.S. analyzed data; P.S. and A.P. wrote the paper.}

\bibliographystyle{jfm}
\bibliography{jfm-instructions}

\end{document}